\documentclass[fleqn,usenatbib,useAMS]{mnras}
\usepackage[utf8]{inputenc}
\usepackage{graphicx, natbib, color, bm, amsmath, epsfig, amstext, array, hhline}

\usepackage[T1]{fontenc}
\usepackage{ae,aecompl}
\usepackage{newtxtext,newtxmath}

\newcommand{\M}{M_{\text{1D}}}
\newcommand{\tdyn}{\tau_{\text{dyn.}}}

\newcommand{\D}{\mathrm{d}}

\newcommand{\mach}{\ensuremath{m}}

\title{Finite Shock Model of Density Fluctuations in Isothermal Turbulence}

\title{Finite Shock Model of Density in Supersonic Turbulence}

\author[B. Rabatin et al.]{Branislav Rabatin,$^{1}$%
\thanks{Contact e-mail: \href{mailto:br18b@fsu.edu}{br18b@fsu.edu}}%
David C. Collins $^{1}$
\\
$^{1}$Florida State University, Tallahassee, FL 32309}

\begin{document}
\label{firstpage}
\pagerange{\pageref{firstpage}--\pageref{lastpage}}
\maketitle

\begin{abstract}
    The probability distribution of density in isothermal, supersonic, turbulent gas is approximately lognormal. This behaviour can be traced back to the shock waves travelling through the medium, which randomly adjust the density by a random factor of the local sonic Mach number squared. Provided a certain parcel of gas experiences a large number of shocks, due to the central limit theorem, the resulting distribution for density is lognormal. We explore a model in which parcels of gas undergo finite number of shocks before relaxing to the ambient density, causing the distribution for density to deviate from a lognormal. We confront this model with numerical simulations with various r.m.s. Mach numbers ranging from subsonic as low as 0.1 to supersonic at 25. We find that the fits to the finite formula are an order of magnitude better than a lognormal. The model naturally extends even to subsonic flows, where no shocks exist.
    
\end{abstract}

\section{Introduction}

Density and its statistics play crucial role in the dynamics of molecular clouds and star formation, the centerpiece of astrophysical processes \citep{BigProblems, PadoanSFR14}. Isothermal turbulence is ubiquitous in modeling astrophysical settings, as it is a relatively accurate model of efficiently cooled molecular clouds, capable of explaining the observed density fluctuations \citep{Elmegreen04}. Observations suggest that supersonic turbulence is dominant in star-forming regions \citep{Scalo04}. Numerical models have shown that supersonic turbulence, while inhibiting the collapse by increasing the effective Jeans mass, also gives rise to large density variations within the medium, allowing for a local collapse \citep{MacLow04}.

The probability distribution function (PDF) of the fluid density, $\rho$, is usually treated as lognormal due to the self-similar statistics within turbulent isothermal medium \citep{Vazquez-Semadeni94, PNJ97}.  Letting $s=\ln \rho/\rho_0$, the PDF of $s$ can be written as
\begin{equation}
	f_s (s; \sigma) \, \D s = \frac{1}{\sqrt{2 \pi \sigma^2}} \exp \left( - \frac{\left( s -\mu \right)^2}{2 \sigma^2} \right) \D s,
\end{equation}
where the mean value $\mu \equiv \left\langle s \right\rangle = - \sigma^2/2$ ensures $\langle e^s \rangle = 1$. The variance $\sigma^2 \equiv \left\langle s^2 \right\rangle - \left\langle s \right\rangle^2$ depends on the r.m.s. 1D sonic Mach number, $\M$, and the ratio of rotational forcing to compression forcing, $\xi$  \citep{Passot98, Federrath08, Schmidt09}.
It has been found that $\sigma^2 = \log \left( 1 + b^2 \text{M}_{\text{3D}}^2 \right)$ \citep{Passot98, Federrath08, Federrath10}, where $b=1/3$ for purely rotational forcing ($\xi=1$) and $b=1$ for compressive forcing ($\xi=0$).
The explicit form of the PDF of density can be used to understand many astrophysical phenomena, such as star formation \citep{Krumholz05, Padoan11, Hennebelle11, Federrath12}, mass distribution \citep{Padoan02} and chemical evolution \citep{Pringle01, Gaches15}.

A lognormal distribution emerges from a large number of independent multiplicative events, to which, upon taking the logarithm of the random variable, the central limit theorem applies. If we treat supersonic isothermal turbulence as an ensemble of shocks, the properties of a parcel of gas are adjusted by the shock jump conditions \citep{Rankine, Hugoniot1, Hugoniot2}, that, for density, result in
\begin{equation}
	\frac{\rho_2}{\rho_1} = \mach^2
	\label{eq_density_jump}
\end{equation}
where $\rho_1$ and $\rho_2 > \rho_1$ are pre- and post-shock densities, respectively, and $\mach$ is the upstream local sonic Mach number. Velocity and density are usually treated as independent random variables \citep[e.g.][]{Kritsuk07, Federrath10, Pijpers97}. Thus, the log of density can be treated as a large number of additive events, and the distribution of the log of density is expected to be roughly Gaussian.

In reality, a parcel of gas does not merely experience a large number shock jumps. It also contains rarefaction waves where the pressure caused by the shock pushes the density  back to the mean density. Therefore, gas observed after a long time of turbulent driving, cannot be regarded as having experienced infinitely many shock jumps, as the density eventually resets to the ambient value. Shocks and rarefactions are in constant opposition, and the parcel has a memory of a finite number of shocks.

In this work we consider a simple model assuming that a parcel of gas undergoes a finite number of shocks, $n$, before relaxing to the background density. The local sonic Mach number $\mach$ is drawn from a Maxwell-Boltzmann distribution with a global Mach number $\M$. Then we compute the PDF of $s$ for $n$ such shocks.
For low values of $n$, this results in a pronounced tail for low densities, while the peak of the distribution shifts towards high density.  We perform a suite of simulations with 1d r.m.s. Mach number ranging from 0.1 to 25, and a variety of forcing parameters, $\xi$, and show that the error on the fit to the distribution is as much as an order of magnitude
lower than the fit to a plain lognormal. We also derive a prediction for the number of shocks experienced by a parcel of gas based only on $\mu$ and $\sigma$, which are easily computed from numerical data.

Other works that model non-lognormality of the density statistics can be divided into two classes. In the first class, additional phenomena, such as gravity \citep{Klessen_gravity20, Slyz05, Collins11, Kritsuk_gravity_2011, Federrath_Klessen_gravity_2013, Girichidis_gravity_2014} or more complicated thermodynamics \citep{Nolan_thermo_15, Federrath_thermo_15, Scalo98} are found to change the distribution of density, adding power law wings to the high- or low-density end. Other works focus on explaining the anomalous density fluctuations purely within the turbulent framework. \citet{Mocz19} explore a Markov process with variable drift and diffusion timescales, giving rise to a PDF with a steeper high-density tail, while the low-density end remains unchanged. \citet{Squire17} propose a compound log-Poisson process which also produces a tilted log density PDF. A model of quantized log-Poisson cascades leading to a PDF with weight redistribution towards low densities is considered by \citet{Hopkins13}.

In Section \ref{sec.theory}, we derive the PDF of $s = \log \rho / \rho_0$ for $n$ shocks, $f(s; n)$, and our prediction for the number of shocks given the moments of $s$. In Section \ref{sec.results} we compare our results with simulation. The summary of our results can be found in \ref{sec.conclusions}.
\section{Density fluctuations}
\label{sec.theory}

If we consider the density within a certain parcel of gas to be a product of infinitely random factors, then, according to the central limit theorem, such product results in the lognormal distribution.
To explore the case when the number of shocks $n$ that a particular portion of the gas experiences is finite, we make use of the characteristic function $\phi (s; n)$ of the resulting random variable, $s$. In our specific case of a uniform, isotropic forcing with a global sonic Mach number $\M$, the local Mach number $\mach$ is drawn from a Maxwell distribution
\begin{equation}
	\mach \sim f_v (\mach; \M) = \frac{4 \pi \mach^2}{(2 \pi \M^2)^{3/2}} \exp \left( - \frac{\mach^2}{2 \M^2} \right)
\end{equation}

Without loss of generality, we introduce the variable $Y$, which is normalized to zero mean and unit variance, 
\begin{align}
    Y &= \frac{\log \mach^2 - \mu_{\log \mach^2}}{\Sigma} \label{eqn_Y} \\
    \mu_{\log \mach^2} &= \left\langle \log \mach^2 \right\rangle = \log \left( \frac{\M^2 e^{2 - \gamma}}{2} \right) \label{eqn_mulogM} \\
    \Sigma^2 &= \left\langle \log^2 \mach^2 \right\rangle - \left\langle \log \mach^2 \right\rangle^2 = \frac{\pi^2}{2} - 4 \label{eq_SIGMA}
\end{align}
where $\gamma \approx 0.5772$ is the Euler-Mascheroni constant.

Finally we model $s$ as the sum of $n$ independent, identical events $Y$ as
\begin{equation}
    s = \frac{1}{\sqrt{n}} \sum_{i = 1}^n Y_i,
\end{equation}
where $\sqrt{n}$ is necessary to ensure $\sigma = 1$.

We proceed by finding the characteristic function for $s$,
\begin{multline}
    \phi (\omega; n) \equiv \left\langle e^{i \omega s} \right\rangle = \left\langle e^{i \omega Y / \sqrt{n}} \right\rangle^n = \\
    = \left( 4 e^{\gamma - 2} \right)^{\sqrt{n} i \omega / \Sigma} \left[ \frac{2}{\sqrt{\pi}} \Gamma \left( \frac{3}{2} + \frac{i \omega}{\sqrt{n} \Sigma} \right) \right]^n .
    \label{eq_phi_full}
\end{multline}

The probability distribution for $s$ is given by the inverse Fourier transform of $\phi$
\begin{equation}
	f_s (s; n) = \int \limits_{- \infty}^{\infty} \frac{\D \omega}{2 \pi} e^{- i \omega s} \phi (\omega; n) .
	\label{eq_Fourier}
\end{equation}

This form will be used explicitly when comparing the theoretical PDF to simulated datasets.

In the limit of $n \gg 1$ the characteristic function can be approximated as $\phi (\omega, n) = \left( 1 - \omega^2 / 2n \right)^n \to e^{-\omega^2/2}$ which corresponds to the characteristic function of normal distribution with zero mean and unit variance, $f_s (s; n \to \infty) = (2 \pi)^{-1/2} e^{-s^2/2}$.

Finite values of $n$ give rise to a weight redistribution along the ideal Gaussian shape into a PDF in which the weight is shifted towards lower densities. This also results to a shift of the peak of the distribution, that can be calculated analytically. By considering the lowest order correction to $\phi (\omega; n)$ in $1/n$
\begin{equation}
	\phi (\omega; n) \approx e^{- \omega^2/2} \left[ 1 + \frac{i \alpha}{3} \frac{\omega^3}{\sqrt{n}} \right],
	\label{eq_phi_approx}
\end{equation}
with $\alpha$, equal to $(7 \zeta (3) / 8 - 1)/\Sigma^{3/2} \approx 0.4585$, a constant appearing in virtually all subsequent approximate formulas. Subsequently, by using \eqref{eq_Fourier} with the approximate form \eqref{eq_phi_approx} for $\phi$, $f (s; n)$ expanded to the lowest order in $1/n$ is
\begin{equation}
	f (s; n) \approx \frac{1}{\sqrt{2 \pi}} e^{- s^2/2} \left[ 1 + \frac{\alpha}{3 \sqrt{n}} s \left( s^2 - 3 \right) \right].
	\label{eq_fs_approx}
\end{equation}
To the lowest order in $1/n$, the approximate expression for the mode (maximum of \eqref{eq_fs_approx}) of the distribution is
\begin{equation}
	s_{\text{max}, n} \approx \frac{\alpha}{\sqrt{n}} > 0.
\end{equation}

\begin{figure}
    \begin{center}
        \includegraphics[width=0.45\textwidth]{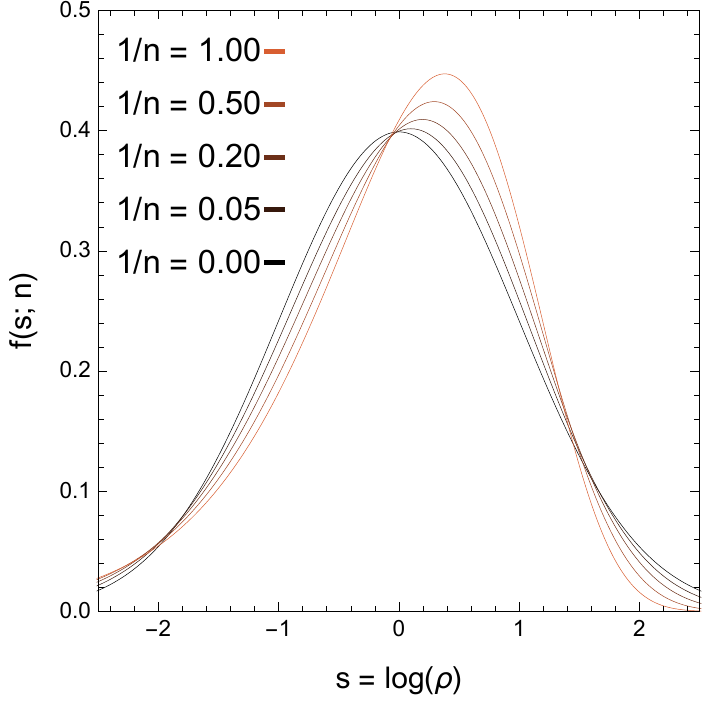}
        \caption[ ]{Plot of $f(s; n)$ calculated numerically for various values of $n$. Note the distinct tilt to the right resulting in a shallowing/steepening of the slope for $s$ below and above the mode.} 
    \label{fig_tilted}
    \end{center}
\end{figure}

Figure \ref{fig_tilted} shows PDFs derived in \eqref{eq_Fourier} using the full form of $\phi$ in \eqref{eq_phi_full} for various $n$. Note the shift in the peak as $n$ becomes finite and the weight redistribution in the tails. The pure lognormal corresponds to $1/n = 0$.

By construction, $f (s; n)$ has unit variance and zero mean. To reintroduce an arbitrary variance and mean of $s$, we shift and rescale the argument as follows
\begin{equation}
	f_s (s;\mu, \sigma, n) = \frac{1}{\sigma} f_s \left( \frac{s - \mu}{\sigma}; n \right)
\end{equation}

The formulation using the characteristic function $\phi$ allows us to compute all central moments of $f (s; n)$, using properties of the Fourier transform
\begin{equation}
	\langle (s-\mu)^k \rangle = \int \limits_{-\infty}^\infty (s-\mu)^k f_s (s; \mu, \sigma, n) = \left( - i \sigma \right)^k \left. \frac{\D^k}{\D \omega^k} \phi(\omega; n) \right|_{\omega = 0}
\end{equation}

Since $\phi (\omega; n)$ is analytic in the whole complex plane except for its singular points occurring at ${\omega_k = i \sqrt{n} (2k+3) / \sigma_{\log m}}$, it is also possible to calculate $\langle e^s \rangle = \langle \rho/\rho_0 \rangle = 1$ to be
\begin{equation}
	\left\langle e^s \right\rangle = 1 = e^\mu \phi (-i \sigma; n).
	\label{eq_rho0_logrho_sigma_n}
\end{equation}
This shows, that the parameters $\rho_0$, $\langle \log \rho \rangle$, $\sigma_{\log \rho}$ and $n$ are not all independent. For large $n$, \eqref{eq_rho0_logrho_sigma_n} can be approximated via expansion in $1/n$ as
\begin{equation}
    1 \approx e^{\mu + \sigma^2 / 2} \left( 1 - \alpha \sigma^3 / 3 \sqrt{n} \right)
\end{equation}
which gives the initial estimate for $n$
\begin{equation}
	n \approx \left[ \frac{\alpha \sigma^3}{3 \left( 1 - e^{-\mu - \sigma^2 / 2} \right)} \right]^2,
	\label{eq_rho0_logrho_sigma_n_approx}
\end{equation}
to serve as the first approximation to $n$ given all other parameters. Note, that for a pure lognormal case, condition $\left\langle e^s \right\rangle = 1$ enforces $\mu = - \sigma^2/2$ in which case the expression for $n$ blows up to infinity, as expected.

\begin{center}
    \begin{table*}
\begin{tabular}{| c | c || c | c | c || c | c | c || c | c || c |}
\hline
$\xi$ & $M_{1D}$ & $-\mu _{\text{est}.}$ & $\sigma _{\text{est}.}$ & $n_{\text{est}.}$ & $-\mu _{\text{fit}}$ & $\sigma _{\text{fit}}$ & $n_{\text{fit}}$ & $\varepsilon _{\ln }\text{($\%$)}$ & $\varepsilon _{\text{fit}}\text{($\%$)}$ & $V_{\text{sh}.}\text{($\%$)}$ \\ \hline
\, & $0.0946$ & $7.89\times 10^{-5}$ & $0.0126$ & $14.8$ & $7.66\times 10^{-5}$ & $0.0124$ & $15.6$ & $7.5\times 10^{-1}$ & $7.0\times 10^{-1}$ & $0$ \\ 
\, & $0.240$ & $2.14\times 10^{-3}$ & $0.0658$ & $4.87$ & $2.25\times 10^{-3}$ & $0.0672$ & $11.3$ & $3.0\times 10^{-2}$ & $2.7\times 10^{-2}$ & $8.60 \times 10^{-3}$ \\ 
\, & $0.474$ & $2.45\times 10^{-2}$ & $0.224$ & $8.82$ & $2.40\times 10^{-2}$ & $0.221$ & $10.4$ & $5.0\times 10^{-3}$ & $3.8\times 10^{-3}$ & $21.1$ \\ 
\, & $0.899$ & $1.52\times 10^{-1}$ & $0.559$ & $33.3$ & $1.53\times 10^{-1}$ & $0.561$ & $31.1$ & $2.6\times 10^{-3}$ & $9.7\times 10^{-4}$ & $75.2$ \\ 
$0.5$ & $1.67$ & $4.79\times 10^{-1}$ & $0.998$ & $59.4$ & $4.83\times 10^{-1}$ & $1.00$ & $62.6$ & $1.0\times 10^{-3}$ & $2.6\times 10^{-4}$ & $95.4$ \\ 
\, & $3.21$ & $9.51\times 10^{-1}$ & $1.43$ & $35.8$ & $9.55\times 10^{-1}$ & $1.44$ & $27.4$ & $5.2\times 10^{-3}$ & $2.6\times 10^{-3}$ & $99.3$ \\ 
\, & $6.38$ & $1.36$ & $1.76$ & $23.6$ & $1.37$ & $1.77$ & $13.8$ & $2.2\times 10^{-3}$ & $5.5\times 10^{-4}$ & $99.9$ \\ 
\, & $12.9$ & $1.57$ & $1.94$ & $17.1$ & $1.58$ & $1.95$ & $7.46$ & $7.1\times 10^{-3}$ & $2.6\times 10^{-3}$ & $99.9$ \\ 
\, & $25.6$ & $1.70$ & $2.08$ & $13.7$ & $1.71$ & $2.06$ & $5.60$ & $1.8\times 10^{-2}$ & $7.9\times 10^{-3}$ & $99.9$ \\ \hline 
$1.$ & $0.359$ & $7.96\times 10^{-3}$ & $0.128$ & $3.11$ & $7.60\times 10^{-3}$ & $0.125$ & $3.23$ & $4.0\times 10^{-2}$ & $3.2\times 10^{-2}$ & $4.37$ \\ 
$0.5$ & $0.325$ & $6.81\times 10^{-3}$ & $0.118$ & $5.57$ & $6.41\times 10^{-3}$ & $0.114$ & $8.22$ & $2.6\times 10^{-3}$ & $2.0\times 10^{-3}$ & $1.73$ \\ 
$0$ & $0.336$ & $2.14\times 10^{-1}$ & $0.677$ & $10.2$ & $2.15\times 10^{-1}$ & $0.679$ & $7.67$ & $2.2\times 10^{-2}$ & $3.1\times 10^{-3}$ & $3.91$ \\ 
$1.$ & $0.986$ & $1.63\times 10^{-1}$ & $0.578$ & $60.7$ & $1.64\times 10^{-1}$ & $0.579$ & $54.4$ & $1.6\times 10^{-3}$ & $7.7\times 10^{-4}$ & $80.8$ \\ 
$0.5$ & $0.994$ & $1.92\times 10^{-1}$ & $0.629$ & $54.8$ & $1.92\times 10^{-1}$ & $0.628$ & $54.2$ & $3.1\times 10^{-3}$ & $9.7\times 10^{-4}$ & $81.3$ \\ 
$0$ & $0.987$ & $1.08$ & $1.60$ & $12.8$ & $1.12$ & $1.64$ & $4.56$ & $6.0\times 10^{-2}$ & $1.6\times 10^{-2}$ & $77.5$ \\ 
$1.$ & $8.48$ & $1.08$ & $1.56$ & $20.5$ & $1.09$ & $1.55$ & $19.5$ & $5.9\times 10^{-3}$ & $3.3\times 10^{-3}$ & $99.9$ \\ 
$0.5$ & $7.99$ & $1.38$ & $1.78$ & $19.9$ & $1.39$ & $1.78$ & $11.6$ & $5.5\times 10^{-3}$ & $1.7\times 10^{-3}$ & $99.9$ \\ 
$0$ & $7.71$ & $4.40$ & $3.76$ & $76.2$ & $4.34$ & $3.80$ & $2.26$ & $9.4\times 10^{-3}$ & $7.2\times 10^{-4}$ & $99.9$ \\
\hline
\end{tabular}
\caption[] {Simulation parameters and fit results. The first column shows the forcing parameter $\xi$, the second column is the 1D Mach number. The next three columns show the parameters of PDF of $s$ estimated from the data. Following three columns feature fitted parameters of the PDF. The penultimate two columns show errors of fits to the data; lognormal and PDF with fitted parameters. Last column shows the volume fraction of the gas that is, on average, capable of undergoing shocks. The first part of the table features various Mach numbers, while the forcing is kept constant. The second part of the table compares three sets of similar Mach numbers at different forcings (purely solenoidal, mixed and compressive).}
\label{tab_results}
\end{table*}
\end{center}

\begin{figure*} \begin{center}
    \includegraphics[width=0.965\textwidth]{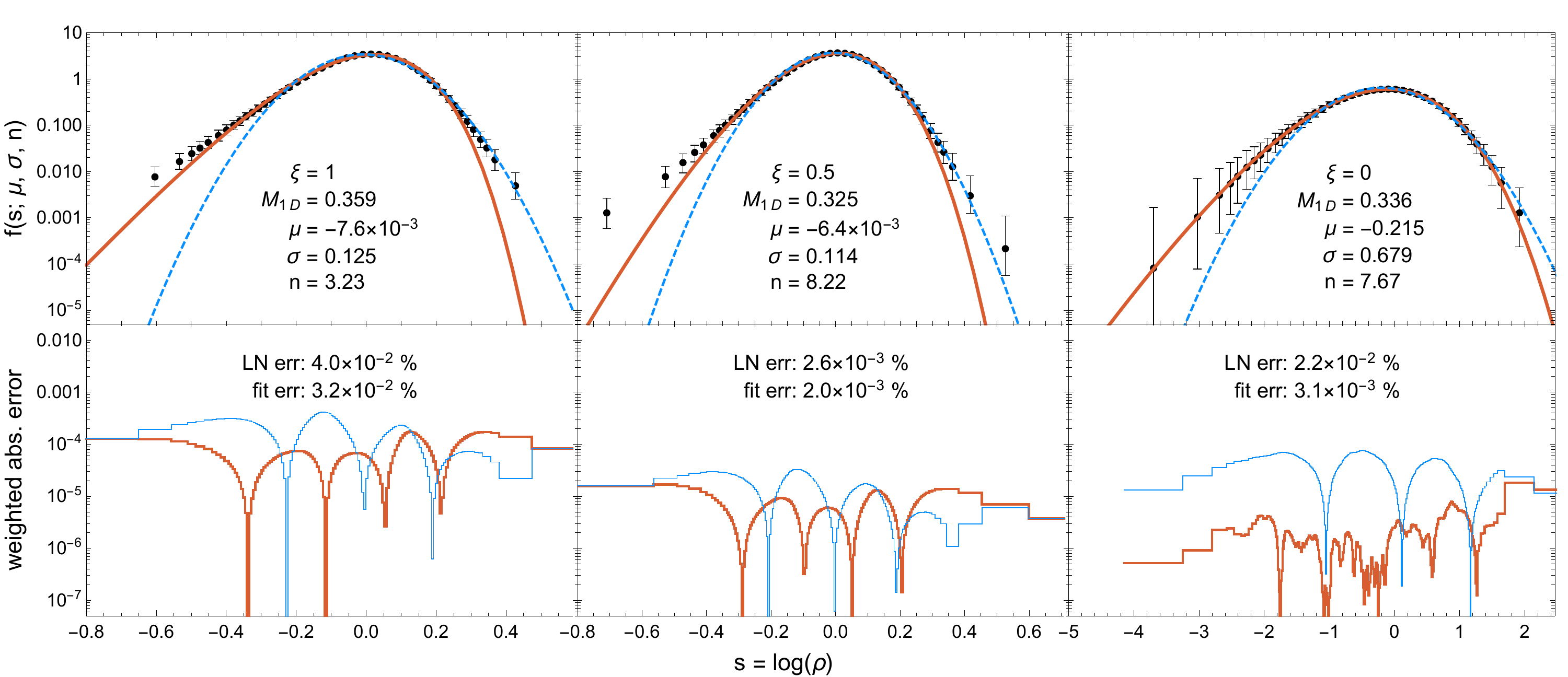}
    \includegraphics[width=0.965\textwidth]{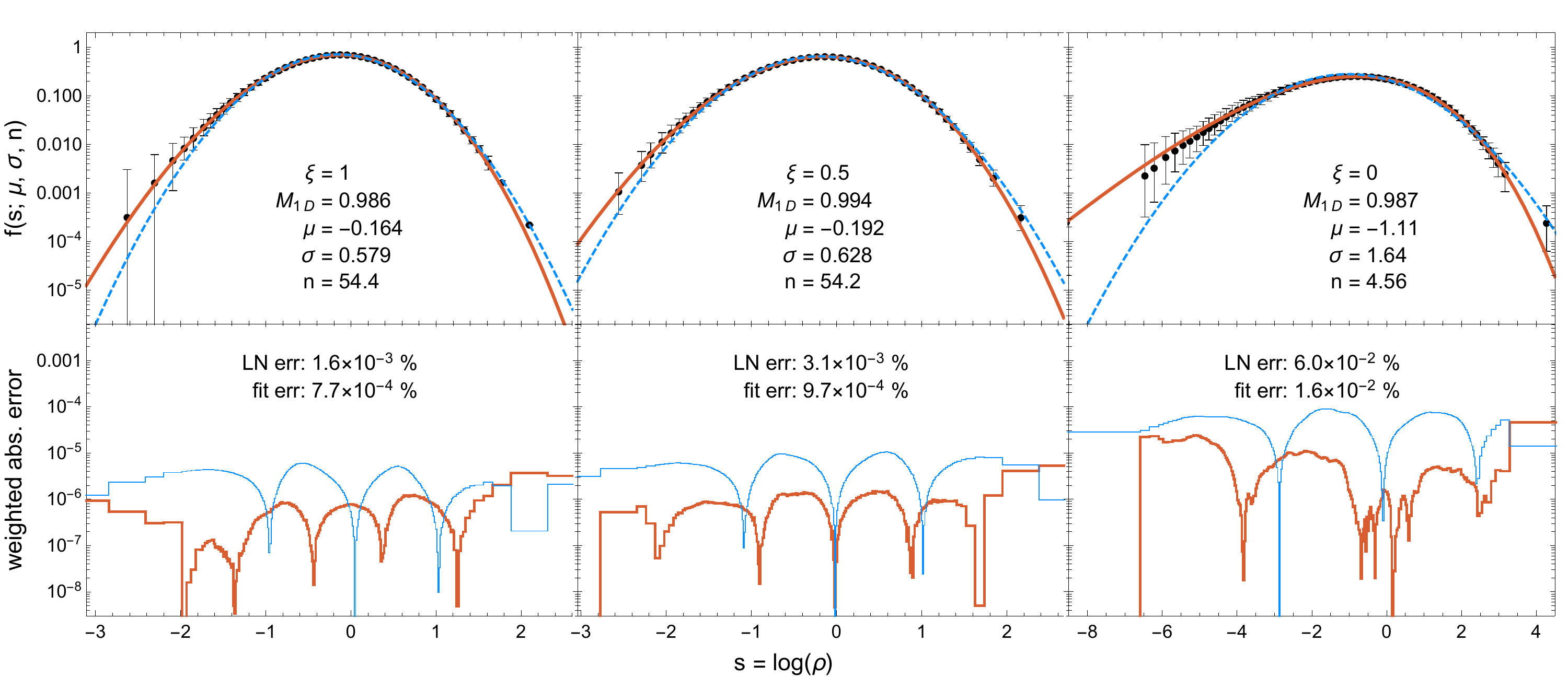}
    \includegraphics[width=0.965\textwidth]{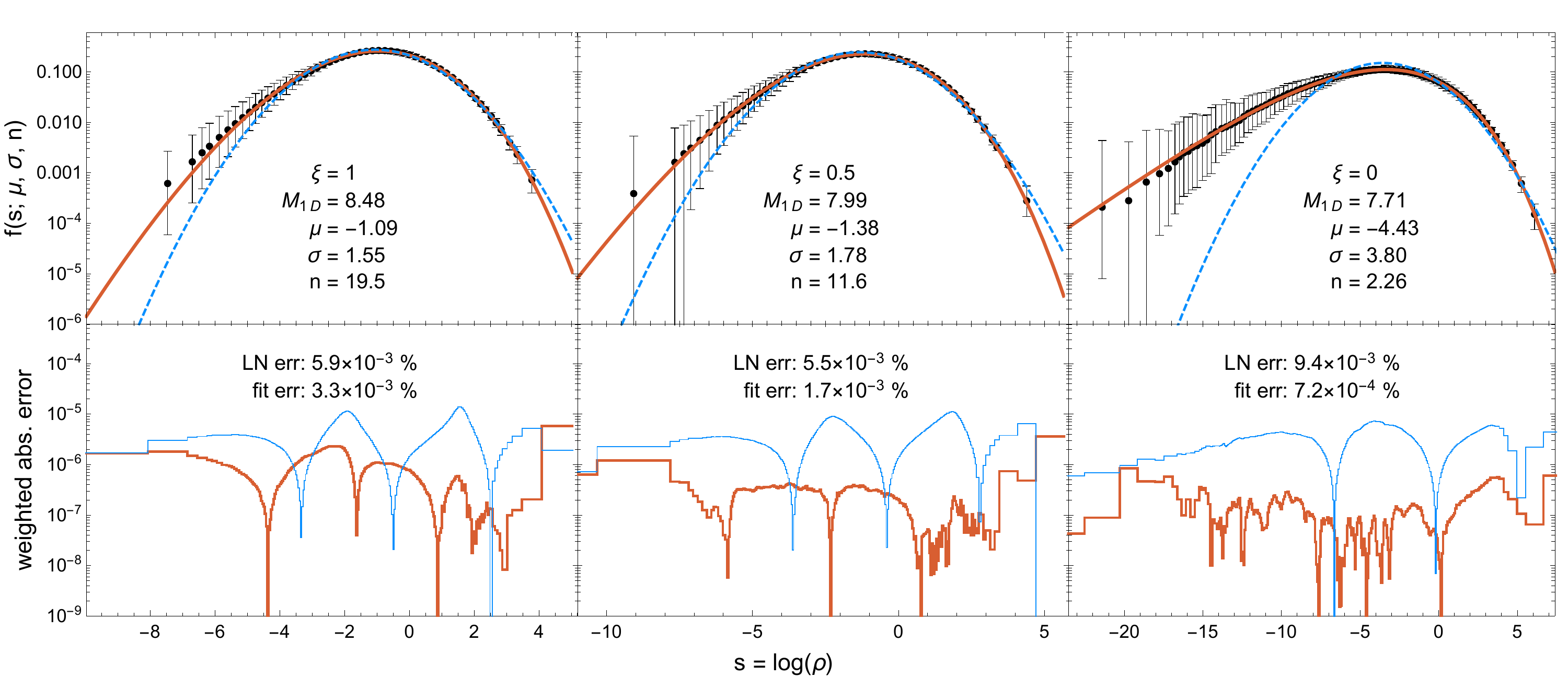}
\caption[ ]{Comparison between the numerical simulations (black points) and our theoretical fit (orange line). The best fit of the lognormal form (blue line) is shown for reference. The overall errors are obtained using eq. \eqref{eq_epsilon}.} 
\label{fig_results} \end{center} \end{figure*}
\section{Confrontation with data}
\label{sec.results}

The theoretical model in the previous section was confronted with numerical data simulated by the open source code Enzo (\cite{Bryan14}) using the piecewise parabolic method (\cite{Woodward84}) at a fixed resolution of $256^3$ cells. Simulations were driven using Stochastic forcing module implemented within Enzo \citep{Schmidt09}. We tested our model on various Mach numbers ranging from subsonic (0.1) to supersonic (25.6) while keeping the ratio of solenoidal to compressive forcing components, $\xi = 0.5$, constant. We then compared different forcing modes ($\xi = 
0, 0.5, 1$) while keeping the Mach number consistent. For each $\M$ we consider the dynamical time $\tdyn$ as the time scale at which two frames become statistically uncorrelated. The dynamical time is roughly equal to the crossing time $\tau_{\text{crossing}} = (L/2)/\sqrt{\langle v^2/3 \rangle}$, or, in code units $1/(2 \M)$. For statistical purposes, only frames with $t \geq 2 \tdyn$ are considered, as the fluid becomes settled in its stochastic turbulent motion. The histograms of $\log \rho$ is obtained averaging frames from $2 \tdyn$ to $30 \tdyn$. These can be seen as the black dots in odd rows in Figure \ref{fig_results}.

Figure \ref{fig_results} visually compares the PDF of $s$ in the odd rows; the black dots indicate geometrical centers of the histograms of the simulated data, blue dashed line is the lognormal fit with a free parameter $\sigma$, while the orange line follows our finite shock fit with the parameters listed on each plot. The even rows show the weighted absolute error between the data and our fit (black line) and lognormal fit (blue line) $\left| f_{\text{data}} (s) - f_{\text{fit}} (s) \right|/\sigma_{\text{data}} (s)$.

The variance in the histograms between frames is usually overwhelmed by the values above the mean probability, so a more nuanced approach allows us to display asymmetric error bars in both directions. The asymmetric variance from the mean probability density fraction $f_b$ within a certain bin $b$ is calculated for both positive and negative direction for all bins within the histogram using our ensemble of 281 frames,
\begin{equation}
    \sigma^{\pm} (\text{$s \in$ bin $b$}) = \left[ \frac{\sum_{\text{frame $i$}} \left( f_{bi} - f_b \right)^2 \theta \left( \pm f_{bi} \mp f_b \right)}{\sum_{\text{frame $i$}} \theta \left( \pm f_{bi} \mp f_b \right)} \right]^{1/2}.
\end{equation}

$\sigma^-$ defined this way is guaranteed to be bounded above by the probability density of the bin. For the purpose of normalizing the absolute difference by $\sigma_{\text{data}} (b)$ in a certain bin $b$ we take the geometric mean of $\sigma^+$ and $\sigma^-$, $\sigma_{\text{data}} (b) = \sqrt{\sigma^+_b \sigma^-_b}$.


In Table \ref{tab_results}, the initial estimates for $\mu_{\text{est.}}$, $\sigma_{\text{est.}}$ computed as the ensemble averages, and $n_{\text{est.}}$, using eq. \eqref{eq_rho0_logrho_sigma_n_approx} (left part of the table) were improved on by fitting to the numerical data producing fits $\mu_{\text{fit}}, \sigma_{\text{fit}}, n_{\text{fit}}$ (middle part of the table). $\sigma_{\text{fit}}$ and $n_{\text{fit}}$ were fitted simultaneously, while \eqref{eq_rho0_logrho_sigma_n} was used to fix $\mu_{\text{fit}}$ and thus preserve $\left\langle \rho \right\rangle$.


In the last three columns of the table we compare the absolute error between various fits weighted by the standard deviation within each bin and data histograms. The weighing is introduced as to give preference to bins with lower statistical noise.
\begin{equation}
    \varepsilon = \frac{1}{W} \sum_{\text{bins} \, b} \frac{|b|}{\sigma_b} \left| f_{\text{data}} (b) - f_{\text{fit}} (b) \right|, \quad W = \sum_{\text{bins} \, b} \frac{|b|}{\sigma_b}
    \label{eq_epsilon}
\end{equation}
The errors between theoretical functions and data are shown in the fourth section of Table \ref{tab_results}; lognormal $\varepsilon_{\text{ln}}$ and estimate improved by fitting to the numerical data $\varepsilon_{\text{fit}}$, respectively. Our finite shock model improves over the simple lognormal fit by up to an order of magnitude for each simulation.

It should be noted, that the description via cascade of shocks only works well in cases with substantial sonic Mach number, as the gas needs supersonic speeds in order to form shocks. The exact portion of the gas, by volume, capable of shocking in each dataset is shown in the last column of the table. Assuming a simple Maxwellian distribution of speeds, less than $1\%$ of the volume of gas with Mach number below 0.3 is moving at supersonic speeds. Therefore, the possible match with the density PDF in the subsonic runs is purely formal.

\section{Conclusions}
\label{sec.conclusions}

In this work we model the PDF of density in isothermal turbulence assuming the number of shocks experienced by a certain parcel of gas, $n$ is fixed and finite, as opposed to infinite. As a result we derive a PDF that slightly deviates from lognormal by weighted tails and mode shifted towards higher densities.  This can be though of as the number of shocks a parcel of gas has a "memory" of, as the enhanced postshock pressure pushes the density back towards the mean. 

We confront the newly derived shape with numerical simulations and find it matches the data much better than a simple lognormal fit. With the exception of the lowest Mach number the weighted absolute error between the data and analytic form stays consistently low even in highly supersonic flows regardless of forcing. For supersonic flows, the finite shock model estimates the data with up to an order of magnitude smaller error compared to the lognormal fit. It should be noted, that the mere estimate of $n$ from \eqref{eq_rho0_logrho_sigma_n_approx}, together with $\mu$ and $\sigma$ that can be simply calculated from the data, give much better fit than the lognormal distribution, fitted or otherwise. Even though the shocks leading to \eqref{eq_density_jump} are only present in a medium with sufficient portion experiencing supersonic speeds, the theory is formally capable of describing density distribution in a subsonic turbulent medium, albeit, with higher error.

Focusing on the supersonic flows, the effect of a finite number of shocks is pronounced in simulations with high Mach number over trans sonic flows. A shock wave passing through the medium compresses the material by a factor of $\mach^2$, therefore, higher Mach numbers, on average, lead to higher densities of shock waves. The total mass conservation, however, necessarily limits the volume available to such shock wave, in turn, limiting the longitudinal size of said shock wave. On the other hand, the rarefaction wave following the shock wave adjusts the density of the region behind the shock towards the mean. Since the shock waves are faster, more frequent and limited in size in more turbulent media, on average, a parcel of gas gets to experience fewer shocks before it resets to the ambient density.

We find that, for fixed supersonic Mach number, compressive forcing ($\xi=0$) results in far fewer shocks that rotational forcing ($\xi=1$). Since the density increase is more pronounced, even at moderate Mach numbers, the increased pressure is larger, and rarefactions will occur more quickly. Thus in compressive forcing, a typical parcel of gas has a memory of only a few shocks.


For subsonic flows, we see the opposite trend; datasets with higher Mach numbers show higher number of shocks than those with low Mach numbers, at least, if estimated from the dataset parameters. This might be due to the smaller total volume available to shock, as the probability of $v > c_s$, given by the tail of the Maxwellian distribution, shrinks.

Our results show that a model of the density PDF that includes a finite number of shocks matches simulated distributions better than a lognormal, which assumes an infinite number of shocks.
\section*{Acknowledgements}

The authors wish to thank the reviewer for insightful comments that improved the work.  Support for this work was provided in part by the National Science Foundation
under Grant AAG-1616026.  Simulations were performed on \emph{Stampede2}, part of the Extreme Science and Engineering Discovery Environment \citep[XSEDE;][]{Towns14}, which is supported by National Science Foundation grant number ACI-1548562, under XSEDE allocation TG-AST140008. 

\section*{Data Availability}

The PDF data used in this article can be found at https://github.com/br18b/Finite-Shock-Model.  Raw simulation data is available upon request (br18b@fsu.edu).
\vspace{-2mm}
\bibliographystyle{mnras}
\bibliography{main.bib}

\bsp	
\label{lastpage}
\end{document}